%% file: main.tex
\documentclass[a4paper,fleqn]{cas-sc}

\usepackage[authoryear]{natbib}
\usepackage{lipsum}
\usepackage{longtable}
\usepackage{graphicx}
\usepackage{subcaption}
\usepackage{array,booktabs}

\def\tsc#1{\csdef{#1}{\textsc{\lowercase{#1}}\xspace}}
\tsc{WGM}
\tsc{QE}

\begin{document}
\let\WriteBookmarks\relax
\def\floatpagepagefraction{1}
\def\textpagefraction{.001}

\shorttitle{Scaffolding Enhance SRL}    

\shortauthors{Anonymous}  

\title [mode = title]{Toward Scalable Co‑located Practical Learning: Assisting with Computer Vision and Multimodal Analytics}  

\tnotemark[1] 


%

\author[1]{Xinyu Li}[orcid=0000-0003-2681-4451]\ead{xinyu.li1@monash.edu}
\author[1]{Linxuan Zhao}[orcid=0000-0001-5564-0185]\ead{linxuan.zhao@monash.edu}
\author[1]{Yueqiao Jin}[orcid=0000-xxxxxxx]\ead{ariel.jin@monash.edu}
\author[1]{Yuchen Liu}[orcid=0000-xxxxxxx]\ead{yuchen.liu1@monash.edu}
\author[2]{Jin Zhou}[orcid=0000-xxxxxx]\ead{jenny.zhou@monash.edu}
\author[1]{Roberto Martinez-Maldonado}[orcid=0000-0002-8375-1816]\ead{roberto.martinezmaldonado@monash.edu}
\author[4,1]{Dragan Gašević}[orcid=0000-0001-9265-1908]\ead{dgasevic@hku.hk}
\author[3,1]{Lixiang Yan}[orcid=0000-0003-3818-045X]\ead{lixiang.yan@monash.edu}

\address[1]{Centre for Learning Analytics at Monash, Monash University, Melbourne, Australia}
\address[2]{Department of Civil and Environmental Engineering, Monash University, Melbourne, Australia}
\address[3]{School of Education, Tsinghua University, Beijing, China}
\address[4]{Faculty of Education and School of Computing \& Data Science, The University of Hong Kong, Hong Kong}





\begin{abstract}
Co-located practical learning leaves evidence in visible actions around patients, task resources and room zones, but these traces are often recovered through live observation or retrospective video review. Fixed wide-angle video could reduce sensing burden, yet a debriefing pipeline must do more than detect behaviours: it must maintain detection after small camera-position shifts, relate the detector-derived behaviour trace to instructor-labelled outcomes and preserve room-zone context. This study evaluates a fixed-camera pipeline in repeated nursing simulation. Using a harmonised six-code taxonomy, we tested YOLO26 target-only training and two-stage source-to-target adaptation across two same-room side-view data sources. We then converted detections from 51 instructor-labelled sessions into one-second behaviour and behaviour-zone traces for rate, ordered-network, transition-network and sequence analyses.

Two-stage adaptation improved mean mAP50 from 0.815 to 0.848 for the 2021 target view and from 0.690 to 0.855 for the smaller 2022 target view; with a balanced target quota of \(N = 22\), the 2022 model reached 0.850 mAP50. In the detector-derived behaviour trace analyses, higher phone use characterised low task-performance sessions. Zone labels changed the interpretation of patient interaction: primary patient-care-zone interaction was stronger in higher-performance sessions, while secondary-zone interaction was stronger in lower-performance sessions. Ordered and transition network models showed that ordered room-zone relations contributed beyond behaviour frequency, with the strongest task-performance classifier using zoned and co-presence features. The resulting trace is most appropriate for searchable simulation debriefing, where instructors inspect detected moments rather than receive automated assessment scores.
\end{abstract}




\maketitle

\input{sections/01-introduction}
\input{sections/02-literature}
\input{sections/03_cv_approach}

\input{sections/04-methods}
\input{sections/05-results}
\input{sections/06-discussion}
\input{sections/07-conclusion}









\printcredits

\bibliographystyle{cas-model2-names}

\bibliography{reference}

\bio{}
\endbio

\bio{}
\endbio

\end{document}

%% file: sections/01-introduction.tex
\section{Introduction}

Learning analytics is easiest to build when the learning environment already records the work. Learning-management systems and online assessment platforms store clicks, submissions, discussion posts and video-playback events with timestamps and activity identifiers, so researchers can model engagement, self-regulation, help seeking and course progress without first creating a new observation system \citep{lang_what_2022, solar2021learning}. The same assumption does not hold in rooms where students learn by moving, handling equipment and responding to events. In laboratories, maker spaces, clinical simulations, design studios and safety-training rooms, learners demonstrate understanding through movement, tool use, physical coordination and responses to task events \citep{timmis2016rethinking, worsley2016situating, martinez2018physical}. Much of the evidence remains visible to people in the room but absent from the data record.

In co-located practical learning, the missing record is not a minor technical inconvenience. Educational evidence is spread across people, objects, task zones and scenario prompts. A learner may show competence by approaching the correct patient, consulting a chart, preparing medication equipment, returning to the bedside after a cue, or coordinating with peers around a shared resource. These behaviours are central to authentic assessment and debriefing, but instructors often reconstruct them from live observation, handwritten notes or retrospective video review rather than from structured analytics data \citep{gulikers2004five, messick1995standards}. The trace must retain enough behaviour, location and timing information for an instructor to find teachable moments after the scenario.

Nursing simulation makes the trace problem concrete. Students practise clinical judgement, task prioritisation and teamwork while moving between a patient bed, documentation materials, medication or equipment areas, a workstation and a teacher-acted distractor \citep{lateef2010simulation, mariani2016nursing, levett2014systematic}. Instructors later use these episodes to discuss patient assessment, recognition of deterioration, information use, medication preparation, communication and team coordination during debriefing \citep{decker2013standards, fanning2007role}. The difficulty is that the episodes are scattered through a full-room video stream. When four students act at the same time, a relevant return to the bedside, a phone-use interval or a shift away from the patient may be hard to locate without replaying long sections of video.

MMLA studies have addressed similar visibility problems by adding sensors to rooms and bodies. Indoor-positioning systems, audio, physiological sensors, depth cameras and related modalities have been used to study co-located collaboration and classroom activity \citep{ochoa_multimodal_2022, cukurova2020promise}. In routine teaching, these modalities add setup work: tags must be distributed and charged, anchors and cameras must be calibrated, streams must be synchronised, and richer sensing can increase privacy and consent concerns \citep{pardo2014ethical, mallavarapu2022exploring}. A method that works for a short research deployment may still fail as routine infrastructure if teaching staff must rebuild the setup for every cohort.

A fixed wide-angle camera reduces that setup burden. It can capture learners, resources and zones from one classroom-side sensor without asking students to wear devices or interact with additional equipment \citep{bosch2018quantifying, li2023cvpe}. For simulation debriefing, occupancy alone is insufficient. A trace that only shows where students stood cannot tell whether they were consulting information, handling medication equipment, interacting with the patient, using a phone or moving between task areas. Fixed-camera analytics needs both the location of activity and the visible action category.

Three issues have to be tested before such a pipeline can be used across repeated teaching. The first is maintenance. A camera may be moved slightly between years even when the room, task ecology and major zones remain stable. Relabelling every new cohort or camera view is rarely feasible, so labelled data from one same-room view should be tested as support for a later similar view. The second is downstream use. A high mAP50 score indicates bounding-box agreement, but it does not show whether the detector-derived behaviour trace relates to instructor-labelled task or collaboration outcomes. The third is room-zone meaning. Patient interaction at the bedside, patient-related activity in a secondary zone, and movement around resources can have different implications for debriefing even when they receive similar behaviour labels.

We evaluate a YOLO26-based fixed-camera pipeline using two same-room side-view nursing simulation data sources. The analysis asks whether source-to-target adaptation maintains six-code detection performance after a small same-room camera-position shift, whether detector-derived behaviour traces relate to instructor-labelled task-performance and collaboration-performance groups, and whether room-zone labels change the interpretation and predictive value of those traces. The intended output is a searchable behaviour-zone trace for instructor-led simulation debriefing, not an automated assessment system.

%% file: sections/02-literature.tex
\section{Literature Review}

\subsection{Online Versus Co-Located Learning Analytics}

Online learning analytics benefits from a close coupling between activity and instrumentation. The same platforms that deliver content also record clicks, submissions, video playback, discussion posts and assessment events, usually with timestamps and identifiers that can be linked to course structure \citep{lang_what_2022, solar2021learning}. These records allow researchers and instructors to model engagement, progress, help seeking and risk with limited additional sensing infrastructure \citep{paulsen2024learning, ramaswami2023use}. The learning activity happens inside the recording system.

Co-located practical learning breaks that coupling. A student can assess a patient, move toward a medication trolley, consult paper notes, point to a monitor or warn a peer without producing a platform log \citep{timmis2016rethinking, worsley2016situating}. The room contains evidence, but the evidence is distributed across bodies, tools, locations and task timing rather than stored as ready-made digital events \citep{martinez2018physical, chua2019technologies}. Full-session video preserves this activity, but raw video leaves instructors and researchers with the same practical problem: the relevant moments still have to be found, coded and compared.

The problem is especially visible when competence is assessed through performance. Authentic assessment emphasises situated tasks in which learners coordinate knowledge, tools and social interaction under contextual constraints \citep{gulikers2004five, miller1990assessment}. In nursing simulation, clinical judgement becomes visible through patient assessment, recognition of deterioration, medication or equipment handling, documentation, patient-facing communication and team coordination \citep{papp2003clinical, mariani2016nursing}. Similar observation demands appear in laboratories, design studios and emergency or fire-safety training, where instructors interpret movement around apparatus, hazards and shared task zones \citep{shapiro2004simulation, chua2019technologies}. Learning analytics for these settings needs records that are machine-readable without stripping away the room-based structure of the task.

\subsection{MMLA Approaches for Making Physical Activity Observable}

MMLA responds to physical trace scarcity by treating position, speech, gesture, tool interaction and physiological signals as analysable learning records \citep{ochoa_multimodal_2022, worsley2016situating}. Indoor-positioning systems can reconstruct who gathered around which table, how long peers worked near one another and how instructors circulated across a room \citep{yan2023socio, yan2022spatial}. Audio, depth, physiological and interaction traces can add speaking turns, posture, stress indicators or tool-use events, giving researchers a richer account of teamwork than location alone \citep{echeverria2024teamslides, zhao2023mets}.

The same richness creates operational work. Indoor-positioning deployments require wearable tags, anchors, charging routines, calibration and device management. Audio pipelines require diarisation and, in many cases, transcription; physiological data require baselining and artefact handling; multimodal analyses require stream synchronisation and additional privacy safeguards \citep{pardo2014ethical, mallavarapu2022exploring, yan2022scalability}. These requirements are manageable in a short research trial, but repeated teaching creates a different constraint: the system must be maintainable by ordinary teaching infrastructure across cohorts \citep{chua2024striving, martinez2023lessons}. A useful MMLA record for practical learning has to balance semantic detail with a sensing setup that does not burden students or staff.

\subsection{Fixed-Camera Computer Vision for Where-and-What Traces}

Computer vision changes the sensing arrangement by moving the instrumentation from learners' bodies to the room. A fixed wide-angle camera can observe learners, resources and task zones from a single sensor, reducing the need for participant-worn devices \citep{bosch2018quantifying, li2023cvpe}. Vision-based classroom analytics has already been used to estimate instructor movement, classroom proximity and engagement-related visual cues \citep{ahuja2019edusense, sumer2021multimodal}. These systems make physical learning more observable, but many of them still emphasise where people are rather than what task action they are performing.

Object detection can add the missing action layer by producing time-stamped bounded behaviour events from whole-room video. Recent detector families improve speed and localisation, making classroom-scale inference more feasible on ordinary hardware \citep{khanam2024yolov11, huang2025deimv2}. In practical simulation, the useful labels are not generic human poses. Instructors need visible categories such as information handling, medication or equipment interaction, patient interaction, phone use and transitional movement. A detector that identifies people accurately but ignores patient care, clinical resources or information use would be difficult to use in debriefing \citep{khosravi2022explainable, swiecki2022assessment}.

Fixed-camera video also raises a maintenance problem that is less visible in one-off computer-vision demonstrations. Teaching rooms are adjusted over time: cameras may be moved, beds and equipment may shift, lighting may change and cohorts may use the room differently. Even a small camera-position shift can alter occlusion, body scale and the appearance of task resources. Source-to-target adaptation can reduce new annotation by reusing labelled data from an existing view and supplementing it with a smaller labelled target-view sample \citep{zhao2024detrs}. For educational use, mean detection performance is insufficient; per-code performance matters because weak detection of patient interaction or equipment handling would undermine downstream review even if easier classes perform well.

\subsection{From Detection Accuracy to Behaviour-Zone Process Evidence}

mAP50 stops at box agreement. It does not show whether the detector-derived behaviour trace helps instructors understand a scenario. The trace becomes MMLA evidence only when detected behaviours can be related to course constructs, such as task performance, collaboration performance, attention to the patient, response to deterioration or distribution of work across team members \citep{cukurova2020promise, swiecki2022assessment}. That link requires analysing the trace as a learning process rather than as a collection of isolated detections.

Space changes the interpretation of an action. In a nursing simulation room, patient interaction near the bedside is more directly connected to patient-facing care than patient-related activity away from the immediate care area. Information handling at a workstation may indicate documentation or consultation, whereas similar visible activity near the patient may support direct assessment. Medication/equipment interaction near a trolley has different instructional meaning from ambiguous hand movement in a peripheral zone. Spatial classroom analytics treats the room as part of the evidence structure rather than a neutral background \citep{martinez2018physical, yan2022spatial}.

When stable regions can be annotated, each detection can be represented as a behaviour-zone event, linking what was detected with where it occurred \citep{li2023cvpe, yan2022scalability}. This representation matches simulation debriefing questions: what did students do, where was their attention directed, and how did they move between patient care, documentation, resources and distraction? A broad behaviour label can remain detectable in whole-room video, while the zone label supplies task context. The empirical question is whether zone information changes the relationship between detector-derived traces and instructor-labelled outcomes.

Session-level counts provide a first representation, but they remove order. Two teams may show similar amounts of patient interaction while differing in whether they returned to the patient after consulting information, whether medication/equipment handling occurred before or after key cues, or whether phone use appeared during clinically important periods. Network and sequence methods retain parts of this organisation. ENA represents co-occurrence among coded behaviours within a defined window \citep{shaffer2016tutorial}. ONA extends this logic to ordered connections \citep{zhao2023analysing}. TNA focuses on adjacent or near-adjacent transitions, and DTW compares temporal shapes while allowing sessions to unfold at different speeds \citep{sakoe2003dynamic, petitjean2011global}. The present analysis compares behaviour-only traces, behaviour-zone traces and ordered representations to test whether detector-derived traces contain process information beyond frequency and beyond detection accuracy.

%% file: sections/03_cv_approach.tex
\section{Study Context and Research Questions}

This study was conducted in an undergraduate nursing simulation classroom used repeatedly across two academic years. Each simulation session took place in the same purpose-built room, where students worked around a patient bed, workstation, medication/equipment area and a teacher-acted family-member distractor. A side-mounted 180-degree wide-angle camera recorded the room-level activity, capturing the patient-facing care space and surrounding movement areas without requiring students to wear sensors or interact with additional devices.

\begin{figure}[!h]
\centering
\includegraphics[width=\linewidth]{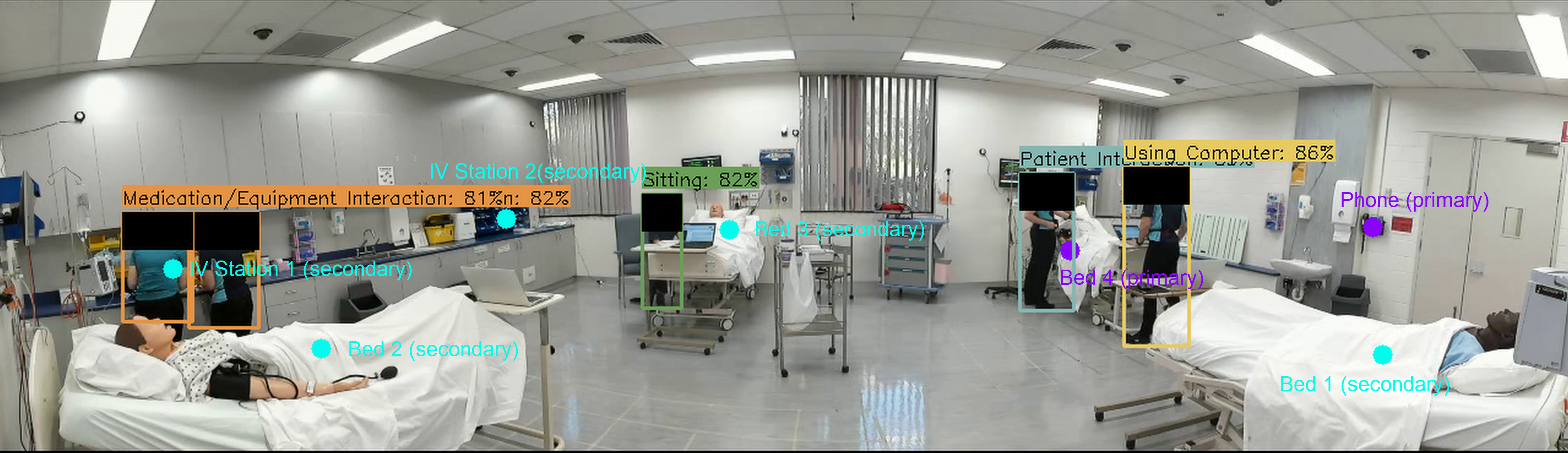}
\caption{Masked wide-angle frame from the nursing simulation room. The overlay shows how detected learner actions were linked to patient-care and secondary room zones for debriefing-oriented trace construction.}
\label{fig:scene_detection}
\end{figure}

The analysis uses two retained same-room side-view data sources, referred to as the 2021 data and the 2022 data. The two archives show the same broad activity ecology: the patient bed area, workstation, medication/equipment area, distractor position and surrounding movement space are all visible from a classroom-side camera position. This pair lets us test whether detector performance can be maintained after a small same-room camera-position shift. A later opposite-side view is excluded from the main analysis because it changes the occlusion structure and the visible arrangement of bodies and furniture; including it would test a stronger cross-view generalisation problem rather than the similar-view maintenance problem addressed here. Figure~\ref{fig:scene_detection} shows the masked camera view, detection overlay and zone annotations used to connect visible behaviours with room location.

\subsection{Instructor-Labelled Session Outcomes}

For the learning-process analyses, the retained video sessions were linked to two instructor-labelled session-level outcomes collected as part of the simulation teaching archive. After each session, facilitating teaching staff completed a six-item formative assessment rubric aligned with the scenario design (Table~\ref{tab:teacher_assessment}). Items T1--T3 assessed task-related performance: structured patient assessment and management, recognition and response to deterioration, and activation of the medical emergency team call. Items T4--T6 assessed collaboration and communication: teamwork, communication of information to the health care team, and understanding of multidisciplinary roles. Each item used a seven-point scale from 1 (\textit{strongly disagree}) to 7 (\textit{strongly agree}).

The task items were summed to produce the task-performance score, and the collaboration items were summed to produce the collaboration-performance score. The archived course-analysis file dichotomised task totals of 12 or higher as high task performance and lower totals as low task performance. Collaboration totals of 18 or higher were coded as high collaboration performance and lower totals as low collaboration performance. These labels are independent of detector training. They are used as instructor-facing session-level outcome contrasts, not as frame-level ground truth for the computer-vision model and not as human-coded action traces.

\begin{longtable}{p{0.12\linewidth}p{0.80\linewidth}}
\caption{Instructor assessment items for task and collaboration performance.}
\label{tab:teacher_assessment}\\
\toprule
Item & Assessment statement \\
\midrule
\endfirsthead
\toprule
Item & Assessment statement \\
\midrule
\endhead
T1 & The students demonstrated a structured approach to patient assessment and management. \\
T2 & The students recognised and responded to early signs of patient deterioration. \\
T3 & The students appropriately activated medical emergency team calls. \\
T4 & The students contributed to effective teamwork. \\
T5 & The students communicated information effectively to the health care team. \\
T6 & The students demonstrated an understanding of the roles of the multidisciplinary team. \\
\bottomrule
\end{longtable}

The final analytic sample contains 51 labelled sessions, with 31 sessions from 2021 and 20 sessions from 2022. The two instructor labels are related but not identical: the sample contains 30 high-task/high-collaboration sessions, 10 high-task/low-collaboration sessions and 11 low-task/low-collaboration sessions, with no low-task/high-collaboration session. This structure reflects the practical teaching archive rather than a balanced experimental design. The low-task group is also numerically concentrated in 2021, with 9 of the 11 low-task sessions from that year. Fisher's exact tests do not reject equal year composition for the task contrast ($p = .166$) or the collaboration contrast ($p = .250$), but the small low-task cell in 2022 limits year-stratified interpretation. The year composition is reported in Supplementary Table S2, and year-stratified replication is treated as a limitation rather than as a separate finding.

\subsection{Ethics and Privacy Protection}

The video archive was collected under Monash University human research ethics approval (Project ID: 28026). Written informed consent was obtained from participating students and teaching staff before data collection. A session was recorded and retained for future research only when all four students in the team had consented to data collection and future research use.

Privacy protection was applied to video examples used in research outputs. Student facial identities were masked by drawing an opaque black rectangle over the upper 20\% of each detected tracking box generated through the OpenCV and YOLO pipeline. This masking procedure protects facial identity in publicly shared research materials and can also be applied in real time if the video stream is used for live review. The present article reports aggregate detector outputs and masked visual examples rather than identifiable student images.

\subsection{Harmonised Six-Code Behaviour Taxonomy}

All detection and learning-process analyses use the harmonised six-code taxonomy shown in Table~\ref{tab:taxonomy}. The original coding scheme separated documentation and computer use, but these labels were merged into \texttt{Information Handling} because both behaviours occur around notes, documentation sheets, laptops and workstation materials in the whole-room side-view recordings. This merge avoids imposing a fine-grained hand-action distinction on video that was collected to capture room-level activity.

The \texttt{Sitting} code indexes the teacher-acted family-member distractor. It is retained in RQ1 because it is part of the visual detection task and provides a control class for the room scene. It is excluded from the learner trace in RQ2 and RQ3 because the learning-process analyses focus on learner-relevant behaviours.

\begin{longtable}{p{0.12\linewidth}p{0.28\linewidth}p{0.52\linewidth}}
\caption{Harmonised six-code behaviour taxonomy.}
\label{tab:taxonomy}\\
\toprule
Code & Behaviour & Interpretation \\
\midrule
\endfirsthead
\toprule
Code & Behaviour & Interpretation \\
\midrule
\endhead
0 & Information Handling & Learner interaction with notes, documentation sheets, laptops, or workstation materials \\
1 & Medication/Equipment Interaction & Learner handling medication trays, trolleys, monitors, pumps, or other clinical resources \\
2 & Other & Learner movement between zones, brief conversation, repositioning, or coordination not assigned to a focal task code \\
3 & Patient Interaction & Learner interaction with the patient, patient bed, bedside monitor, or patient-facing care space \\
4 & Sitting & Teacher-acted family-member distractor; treated as a detection control class and excluded from the learner trace \\
5 & Using Phone & Learner phone use during the scenario \\
\bottomrule
\end{longtable}

\subsection{Research Questions}

The research questions follow the pipeline from detection maintenance to learning-process interpretation. RQ1 asks whether a fixed-camera detector can be maintained when the same simulation room is recorded from a slightly shifted side view. RQ2 asks whether the resulting detector-derived behaviour trace reflects instructor-labelled learning outcomes. RQ3 asks whether adding room-zone information changes the educational meaning and predictive value of the trace.

\textbf{RQ1.} To what extent can source-to-target adaptation maintain six-code YOLO26 detection performance after a small same-room camera-position shift, compared with target-only training?

\textbf{RQ2.} Do YOLO26-derived behaviour traces distinguish instructor-labelled high and low task-performance and collaboration-performance sessions when represented through learner-relevant behaviour counts and temporal process features?

\textbf{RQ3.} Does adding room-zone labels to the detector-derived behaviour trace change how the trace reflects instructor-labelled task-performance and collaboration-performance differences?

%% file: sections/04-methods.tex
\section{Methods}

\subsection{Analytic Design}

The analysis had two parts. The first evaluated detector maintenance across two similar same-room side views. It compared target-only YOLO26 training with a two-stage source-to-target adaptation protocol, using a corrected six-code holdout set for all detection evaluation. This part addresses RQ1.

The second part converted the maintained detector output into detector-derived behaviour traces. The trained detector was applied to the retained 2021 and 2022 session videos, producing one-second behaviour counts. These counts were represented as behaviour-only rates, behaviour-zone rates, ordered networks, transition networks and whole-sequence distance features. These analyses tested whether detector-derived behaviour traces reflected instructor-labelled task-performance and collaboration-performance groups, and whether adding room-zone labels changed the interpretation of those traces. This part addresses RQ2 and RQ3.

\begin{figure}[!h]
\centering
\includegraphics[width=\linewidth]{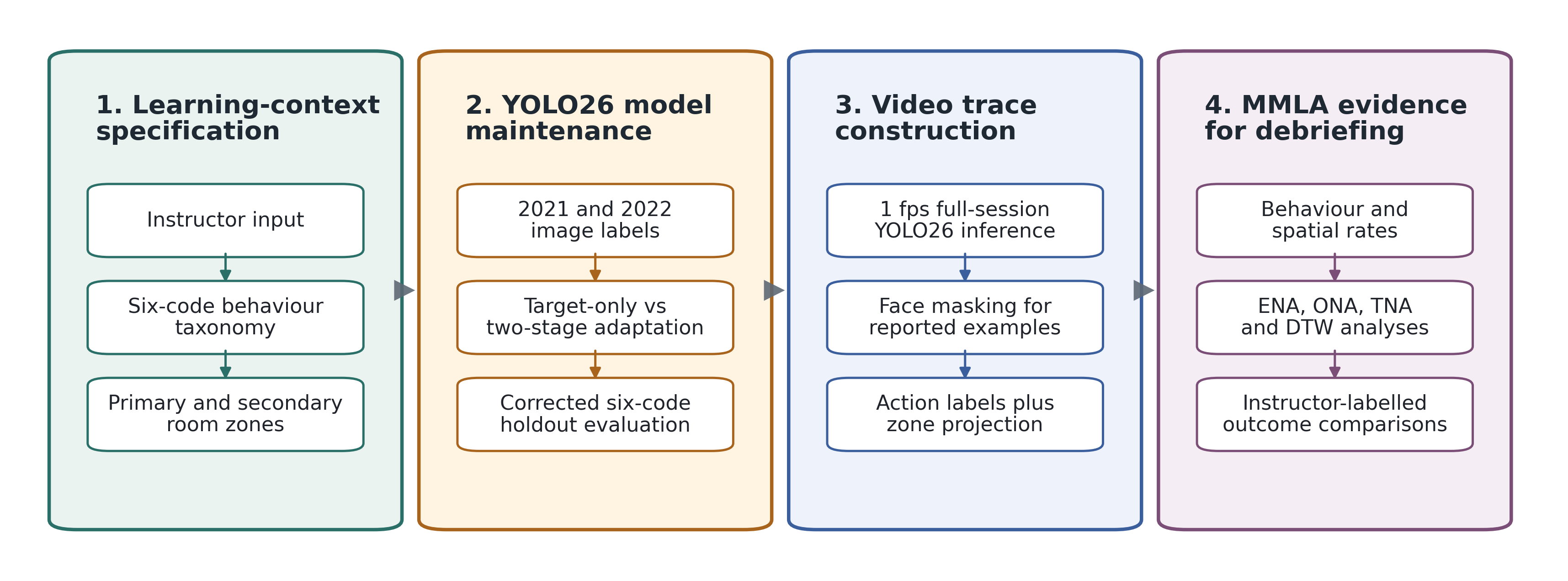}
\caption{Fixed-camera computer-vision and MMLA pipeline. The pipeline converts repeated simulation video into searchable behaviour-zone traces for instructor-led debriefing.}
\label{fig:cv_approach}
\end{figure}

Figure~\ref{fig:cv_approach} summarises the pipeline: define the behaviour codes and room zones, train and evaluate YOLO26 across the two retained camera views, convert full-session video into masked one-second behaviour-zone traces, and analyse those traces against instructor-labelled outcomes.

\subsection{Detection Data and Corrected Holdout Set}

The detection experiments used two retained labelled image pools from the same nursing simulation room: the 2021 data and the 2022 data. Both image pools used the harmonised six-code taxonomy described in Table~\ref{tab:taxonomy}. The 2021 pool contained 791 labelled images, and the 2022 pool contained 225 labelled images. The smaller 2022 pool matches the deployment problem studied here: the later view has less labelled material than the existing source view.

Detection evaluation used a corrected six-code holdout set constructed from the existing labelled archive without adding new image data. The correction merged documentation and computer-use instances into \texttt{Information Handling} and retained phone use as the final six-code index, so that computer-use instances were not inadvertently dropped during taxonomy harmonisation. The corrected holdout set contained 104 images and 463 object instances.

\begin{longtable}{p{0.12\linewidth}p{0.38\linewidth}p{0.30\linewidth}}
\caption{Corrected six-code holdout distribution.}
\label{tab:holdout}\\
\toprule
Code & Behaviour & Instances \\
\midrule
\endfirsthead
\toprule
Code & Behaviour & Instances \\
\midrule
\endhead
0 & Information Handling & 92 \\
1 & Medication/Equipment Interaction & 51 \\
2 & Other & 143 \\
3 & Patient Interaction & 114 \\
4 & Sitting & 47 \\
5 & Using Phone & 16 \\
\textbf{Total} &  & \textbf{463} \\
\bottomrule
\end{longtable}

\subsection{YOLO26 Training and Evaluation}

All detection experiments used YOLO26, a one-stage detector in the YOLO family. The training protocol compared three settings for each target view.

First, the \textit{target-only full} baseline trained directly on the labelled images from the target view and was evaluated on the corrected holdout set. This baseline represents yearly retraining using only the labels available for the camera view being evaluated.

Second, the \textit{two-stage largest-quota} setting first trained on the non-target same-room side view and then adapted on the largest feasible balanced target subset. When the 2021 data were treated as the target view, the 2022 data served as the source view and the largest target quota was \(N = 83\). When the 2022 data were treated as the target view, the 2021 data served as the source view and the largest target quota was \(N = 22\).

Third, the \textit{two-stage full-target} setting used the same source-to-target sequence but adapted on the full target training pool. This setting shows the performance reached when all available target-view labels are used after source-view pretraining.

Each training setting was repeated with seeds 1000 and 4000. The main evaluation metric was mean mAP50 across the six behaviour codes. Per-code mAP50 was also reported because mean performance can hide failure on learner-relevant behaviours. In addition, learner-relevant coverage was calculated as the number of non-\texttt{Sitting} behaviour codes reaching mAP50 \(\geq .70\). \texttt{Sitting} was retained in RQ1 as a detection control class but was excluded from the learner trace used in RQ2 and RQ3.

The session-level video trace used for RQ2 and RQ3 was generated with the full-target two-stage YOLO26 Stage-B model from seed 4000 for each target view, using the corresponding \texttt{best.pt} weights. Video inference used confidence threshold \(=.50\), one frame per second, image size \(=640\), and batch size \(=32\). Using a fixed inference model for each target view ensured that all downstream learning-process analyses were based on one consistent detector output rather than on mixed outputs from multiple seeds. The corrected holdout evaluation established image-level detection performance for the six-code taxonomy. The downstream analyses use detector-derived behaviour traces, not independently human-coded full-session behavioural traces.

\subsection{Video Trace Construction}

The trained YOLO26 models were applied to all retained 2021 and 2022 videos. The resulting detector-derived behaviour trace contained 74,824 one-second rows. After excluding one session without instructor-labelled outcomes, the learning-process analyses used 73,493 one-second rows from 51 labelled sessions. The session counts and one-second rows by outcome group are reported in Supplementary Table S1. In this article, \textit{detector-derived behaviour trace} refers to the one-second YOLO26 output, and \textit{behaviour-zone trace} refers to the same output after zone projection. These traces are treated as process evidence for locating and comparing debriefing-relevant patterns, rather than as validated human-coded behavioural measurement.

Each one-second row contained counts for the six behaviour codes. For RQ2 and RQ3, the learner trace excluded \texttt{Sitting} because that code indexed the seated teacher-acted family-member distractor rather than learner activity. The basic learner trace contained five behaviour codes: \texttt{Information Handling}, \texttt{Medication/Equipment Interaction}, \texttt{Other}, \texttt{Patient Interaction}, and \texttt{Using Phone}.

Because session lengths varied, behaviour counts were converted into session-level rates. For session \(s\) and behaviour code \(c\), the session rate was calculated as

\[
\text{rate}_{s,c} = \frac{\text{count}_{s,c}}{T_s},
\]

where \(T_s\) is the number of one-second rows in session \(s\). These rates were used for the behaviour-only comparisons in RQ2.

\subsection{Zone Annotation and Behaviour-Zone Features}

RQ3 added room-zone context to the detector-derived behaviour trace. Key regions in each camera view were annotated as fixed bounding boxes and then collapsed into two analytically stable regions: a primary task zone and a secondary zone. The primary zone captured the patient-facing task space around the bed and immediate care area. The secondary zone captured the remaining visible room area, including workstation activity, medication/equipment access, peripheral interaction and movement outside the immediate patient-care space. Supplementary Table S3 reports the zone definitions used for behaviour-zone feature construction.

For each YOLO26 detection, the centre point of the predicted bounding box was projected into the annotated room layout. The detection was assigned to the zone whose bounding box contained that centre point. Detections outside both annotated zones were not used in the behaviour-zone features. This procedure produced behaviour-zone labels such as \texttt{primary\_Patient Interaction}, \texttt{secondary\_Patient Interaction}, \texttt{primary\_Information Handling}, and \texttt{secondary\_Medication/Equipment Interaction}.

Two behaviour-zone feature sets were constructed. The first, \textit{zoned learner}, split each learner-relevant behaviour into primary-zone and secondary-zone versions. The second, \textit{zoned plus co-presence}, added four co-presence variables: \texttt{primary\_ind}, \texttt{primary\_col}, \texttt{secondary\_ind}, and \texttt{secondary\_col}. The \texttt{ind} variables indicate individual learner-action presence in a zone during a one-second interval, whereas the \texttt{col} variables indicate that more than one learner-action instance was detected in that zone during the same interval. These variables are interpreted as zone-based co-presence cues, not as direct measures of collaboration quality.

The behaviour-zone representation is coarse by design. The camera provides a whole-room wide-angle view, not a close-up view of hands, devices or speech. Zone annotation separates patient-care-zone activity from activity elsewhere in the room; it does not infer fine-grained clinical procedure quality.

\subsection{Session-Rate Comparisons}

For each outcome, sessions were divided into instructor-labelled high and low groups. The two outcomes were analysed separately as task performance and collaboration performance. Group differences were first examined using session-level rates from the basic learner trace and the behaviour-zone trace.

Cohen's \(d\) was calculated as the standardised difference between the high group and the low group. Positive values indicate higher rates in the high-performing group; negative values indicate higher rates in the low-performing group. Mann--Whitney tests were used as non-parametric group comparisons. Benjamini--Hochberg false-discovery-rate correction was applied within each outcome and feature family. Bootstrap 95\% confidence intervals were calculated with 1,000 session-level resamples for the large effects that were interpreted in the Results and for all reported classification AUCs.

\subsection{Network, Sequence and Classification Analyses}

The one-second traces were also analysed with network and sequence representations to test whether temporal structure added information beyond behaviour frequency.

Epistemic network analysis (ENA) was included as an undirected co-occurrence benchmark \citep{shaffer2016tutorial}. Ordered network analysis (ONA) was used to model directed order relations among coded events \citep{zhao2023analysing}. ENA and ONA used a 30-second stanza window, with session as the unit of analysis. Transition network analysis (TNA) was implemented with the R \texttt{tna} package: detector events were converted to ordered action sequences, session and group transition networks were estimated with \texttt{group\_tna}, and group-level transition differences were tested with \texttt{permutation\_test}. Dynamic time warping (DTW) was used as a whole-sequence distance check, allowing temporal alignment when comparing session-level temporal shapes \citep{sakoe2003dynamic,petitjean2011global}.

For each method and feature set, a leave-one-session-out binary classification model predicted the instructor-labelled outcome. Performance was summarised using AUC, balanced accuracy and bootstrap 95\% confidence intervals from session-level resampling of the cross-validated prediction scores. ONA edge-level group differences were summarised with Cohen's \(d\) and Mann--Whitney tests. TNA transition-level group differences used Cohen's \(d\) from session-level transition weights and permutation \(p\) values from the \texttt{tna} package. Benjamini--Hochberg FDR correction was applied within each method \(\times\) feature-set \(\times\) outcome family, using the same high-minus-low sign convention as the session-rate analyses.

ONA visualisations were generated with the R \texttt{ona} package, which uses the ENA/rENA plotting stack for ordered directed networks. Difference plots are interpreted as descriptive summaries of ordered edge patterns, with inferential claims based on the corresponding edge-level statistics and FDR results.

%% file: sections/05-results.tex
\section{Results}

The results are organised around the three research questions. Section~5.1 reports YOLO26 maintenance across the two similar same-room side views. Section~5.2 examines whether detector-derived behaviour traces distinguish instructor-labelled task and collaboration outcomes. Section~5.3 tests whether behaviour-zone rates change the interpretation of those traces. Section~5.4 reports temporal-network and classification results.

\subsection{RQ1: Maintaining YOLO26 Detection Across Similar Side Views}

Table~\ref{tab:yolo_results} reports YOLO26 performance on the corrected six-code holdout, averaged across seeds 1000 and 4000. The target-only full rows show performance when the detector was trained only on labels from the target camera view. The two-stage rows show performance after source-view pretraining followed by target-view adaptation.

\begin{table}[!h]
\centering
\small
\caption{YOLO26 comparison on the corrected six-code holdout, averaged over seeds 1000 and 4000.}
\label{tab:yolo_results}
\resizebox{\textwidth}{!}{%
\begin{tabular}{@{}lllllllll@{}}
\toprule
Target view & Setting & Overall & Information & Medication/Equipment & Other & Patient & Sitting & Phone \\
\midrule
2021 data & Target-only Full & 0.815 & 0.784 & 0.692 & 0.726 & 0.781 & 0.927 & 0.980 \\
2021 data & Two-stage N = 83 & 0.815 & 0.712 & 0.707 & 0.732 & 0.782 & 0.990 & 0.965 \\
2021 data & Two-stage Full & 0.848 & 0.793 & 0.778 & 0.777 & 0.822 & 0.993 & 0.927 \\
2022 data & Target-only Full & 0.690 & 0.609 & 0.518 & 0.671 & 0.634 & 0.915 & 0.790 \\
2022 data & Two-stage N = 22 & 0.850 & 0.833 & 0.743 & 0.747 & 0.816 & 0.976 & 0.987 \\
2022 data & Two-stage Full & 0.855 & 0.847 & 0.755 & 0.795 & 0.784 & 0.967 & 0.980 \\
\bottomrule
\end{tabular}}
\end{table}

Two-stage adaptation improved overall mAP50 in both target views, but the size of the gain differed by year. For the 2021 target view, full two-stage adaptation increased mean mAP50 from 0.815 to 0.848. The largest balanced quota, \(N = 83\), matched the target-only mean of 0.815, while improving some individual classes, including \texttt{Medication/Equipment Interaction} and \texttt{Sitting}. In the 2021 target view, the clearest gain appeared when all target labels were used after source-view pretraining.

The 2022 target view showed a larger maintenance effect. Target-only full training reached 0.690 mAP50, whereas two-stage full adaptation reached 0.855. The largest balanced target quota, \(N = 22\), reached 0.850, within 0.005 of the full two-stage result and 0.160 above the target-only baseline. This indicates that, for the smaller 2022 target archive, a small balanced target set was sufficient to recover almost all of the full-adaptation performance.

Per-code results show that the improvement was not limited to the control class. In the 2022 target view, \texttt{Information Handling} increased from 0.609 to 0.847, \texttt{Medication/Equipment Interaction} from 0.518 to 0.755, \texttt{Other} from 0.671 to 0.795, and \texttt{Patient Interaction} from 0.634 to 0.784. In the 2021 target view, the largest learner-code gain was again for \texttt{Medication/Equipment Interaction}, which increased from 0.692 to 0.778 under full two-stage adaptation. After excluding the \texttt{Sitting} control class, all five learner-relevant behaviours reached mAP50 \(\geq .70\) in the full two-stage setting for both target views.

For RQ1, source-to-target adaptation maintained six-code YOLO26 performance after a small same-room camera-position shift, with the strongest annotation-saving effect in the smaller 2022 target archive.

\subsection{RQ2: Behaviour-Only Traces and Instructor-Labelled Outcomes}

The learning-process analyses used 51 instructor-labelled sessions: 40 high and 11 low task-performance sessions, and 30 high and 21 low collaboration-performance sessions. Because the low task-performance group is small and concentrated mainly in the 2021 data, the group comparisons in RQ2 and RQ3 are interpreted as exploratory detector-derived evidence rather than as validated behavioural assessment. Table~\ref{tab:basic_rates} reports mean detector-derived session rates for the five learner-relevant behaviours, with \texttt{Sitting} excluded. Rates are expressed as detections per one-second interval, so values can exceed 1 when multiple learner-action instances occur in the same second.

\begin{table}[!h]
\centering
\small
\caption{Mean session rates by instructor-labelled group, using the basic learner trace.}
\label{tab:basic_rates}
\resizebox{\textwidth}{!}{%
\begin{tabular}{@{}lllllll@{}}
\toprule
Outcome group & Sessions & Information & Medication/Equipment & Other & Patient & Phone \\
\midrule
Task high & 40 & 0.554 & 0.315 & 1.363 & 0.690 & 0.102 \\
Task low & 11 & 0.558 & 0.335 & 1.384 & 0.671 & 0.151 \\
Collaboration high & 30 & 0.549 & 0.332 & 1.389 & 0.710 & 0.099 \\
Collaboration low & 21 & 0.564 & 0.301 & 1.336 & 0.652 & 0.132 \\
\bottomrule
\end{tabular}}
\end{table}

For task performance, the clearest behaviour-only difference was phone use. Low task-performance sessions had a higher phone-use rate than high task-performance sessions, 0.151 versus 0.102, with a large negative effect because Cohen's \(d\) is defined as high minus low (\(d = -1.06\), \(p = .013\)). In contrast, total patient interaction showed little task-performance separation: high task-performance sessions averaged 0.690 and low task-performance sessions averaged 0.671 (\(d = 0.14\), \(p = .429\)).

For collaboration performance, the high group had a higher total patient-interaction rate than the low group, 0.710 versus 0.652 (\(d = 0.43\), \(p = .045\)). Phone use again pointed in the opposite direction, with higher rates in the low collaboration-performance group, 0.132 versus 0.099 (\(d = -0.67\), \(p = .052\)). These behaviour-only results show that the detector-derived behaviour trace contains instructor-outcome signal, but the signal is uneven: phone use is the most stable behaviour-level contrast, while total patient interaction is less informative without room-zone context.

\subsection{RQ3: Behaviour-Zone Rates}

Table~\ref{tab:spatial_rates} reports the zone split for the main patient-interaction, phone-use and co-presence variables. The behaviour-zone representation changes the interpretation of patient interaction. In the behaviour-only trace, total patient interaction differed little between high and low task-performance sessions. After adding zone labels, the two patient-interaction components had opposite signs.

\begin{table}[!h]
\centering
\small
\caption{Behaviour-zone and co-presence rates by instructor-labelled group.}
\label{tab:spatial_rates}
\resizebox{\textwidth}{!}{%
\begin{tabular}{@{}lllllll@{}}
\toprule
Outcome group & Sessions & Primary patient & Secondary patient & Primary phone & Primary co-presence & Primary individual \\
\midrule
Task high & 40 & 0.558 & 0.132 & 0.099 & 0.483 & 0.312 \\
Task low & 11 & 0.435 & 0.236 & 0.151 & 0.398 & 0.362 \\
Collaboration high & 30 & 0.582 & 0.128 & 0.098 & 0.497 & 0.303 \\
Collaboration low & 21 & 0.459 & 0.193 & 0.128 & 0.419 & 0.351 \\
\bottomrule
\end{tabular}}
\end{table}

For task performance, high-performing sessions had more primary-zone patient interaction than low-performing sessions, 0.558 versus 0.435 (\(d = 1.04\), 95\% CI [0.36, 1.88], \(p = .006\)). Low-performing sessions had more secondary-zone patient interaction, 0.236 versus 0.132 (\(d = -0.94\), 95\% CI [-1.67, -0.15], \(p = .027\)). These opposite-signed effects explain why the aggregate patient-interaction rate in Table~\ref{tab:basic_rates} shows little task-performance separation.

The collaboration-performance contrast showed a similar primary-zone pattern. High collaboration-performance sessions had more primary-zone patient interaction than low collaboration-performance sessions, 0.582 versus 0.459 (\(d = 1.09\), 95\% CI [0.55, 1.79], \(p < .001\)). This effect remained significant after FDR correction across the session-rate feature family (\(q = .013\)). The corresponding task-rate effects remained exploratory after FDR correction.

Co-presence variables added a second zone-based signal. For task performance, primary-zone co-presence was higher in the high group, 0.483 versus 0.398 (\(d = 0.73\), \(p = .029\)), whereas primary-zone individual presence was higher in the low group, 0.362 versus 0.312 (\(d = -0.65\), \(p = .048\)). For collaboration performance, primary-zone co-presence was also higher in the high group (\(d = 0.68\), 95\% CI [0.11, 1.34]). These variables are reported as zone-based co-presence cues only; they do not measure verbal coordination or collaboration quality directly.

\subsection{Temporal-Network and Classification Results}

Table~\ref{tab:classification_results} compares ENA, ONA, TNA and DTW representations. The table reports leave-one-session-out classification performance for each instructor-labelled outcome and the strongest edge or transition effect where applicable. ENA is included as an undirected co-occurrence benchmark. ONA and TNA test whether order and transition structure add information beyond co-occurrence, and DTW provides a less interpretable whole-sequence distance check. The rows should be read as complementary representations of the same detector-derived trace rather than as competing assessment models.

\begin{table}[!h]
\centering
\small
\caption{Main classification rows for instructor-labelled outcomes. AUC CIs are session-bootstrap 95\% intervals. The \(q\) column reports BH-corrected edge or transition \(p\) values within each method \(\times\) feature-set \(\times\) outcome family.}
\label{tab:classification_results}
\resizebox{\textwidth}{!}{%
\begin{tabular}{@{}lllllll@{}}
\toprule
Outcome & Method & Feature set & AUC [95\% CI] & Balanced accuracy & Strongest effect & q / FDR effects \\
\midrule
Task & ENA & Basic learner & 0.673 [0.448, 0.870] & 0.635 & \texttt{phone--other}, d = -1.16 & .043 / 1 \\
Task & ENA & Zoned learner & 0.639 [0.425, 0.831] & 0.552 & \texttt{P:Phone--S:Info}, d = -1.24 & .073 / 0 \\
Task & ONA & Basic learner & 0.700 [0.487, 0.881] & 0.615 & \texttt{phone -> patient}, d = -1.58 & .007 / 1 \\
Task & ONA & Zoned learner & 0.775 [0.574, 0.948] & 0.731 & \texttt{P:Phone -> P:Patient}, d = -1.47 & .039 / 1 \\
Task & ONA & Zoned + co-presence & 0.820 [0.624, 0.974] & 0.801 & \texttt{P:Patient -> primary\_ind}, d = 1.49 & .027 / 3 \\
Task & TNA & Zoned + co-presence & 0.709 [0.532, 0.863] & 0.640 & \texttt{P:Info -> P:Phone}, d = -1.77 & .098 / 0 \\
Task & DTW & Zoned + co-presence & 0.691 [0.481, 0.879] & 0.591 & whole-sequence distance & NA \\
Collaboration & ENA & Basic learner & 0.543 [0.370, 0.715] & 0.505 & \texttt{phone--other}, d = -0.67 & .303 / 0 \\
Collaboration & ENA & Zoned learner & 0.395 [0.237, 0.564] & 0.464 & \texttt{P:Phone--S:Patient}, d = -0.61 & .869 / 0 \\
Collaboration & TNA & Basic learner & 0.690 [0.532, 0.832] & 0.612 & \texttt{information -> phone}, d = -1.14 & .025 / 1 \\
Collaboration & TNA & Zoned learner & 0.765 [0.615, 0.902] & 0.757 & \texttt{S:Info -> P:Phone}, d = -0.90 & .100 / 0 \\
Collaboration & TNA & Zoned + co-presence & 0.702 [0.541, 0.840] & 0.643 & \texttt{secondary\_col -> P:Phone}, d = -1.16 & .196 / 0 \\
Collaboration & ONA & Zoned + co-presence & 0.711 [0.546, 0.854] & 0.733 & \texttt{P:Phone -> primary\_col}, d = -1.24 & .062 / 0 \\
Collaboration & DTW & Zoned + co-presence & 0.683 [0.526, 0.831] & 0.652 & whole-sequence distance & NA \\
\bottomrule
\end{tabular}}
\end{table}

For task performance, ordered behaviour-zone representations performed better than behaviour-only or undirected co-occurrence representations. Basic ONA reached AUC = 0.700, while zoned learner ONA reached AUC = 0.775. Adding co-presence produced the strongest task-performance classifier, ONA with zoned plus co-presence features, with AUC = 0.820, 95\% CI [0.624, 0.974], and balanced accuracy = 0.801. Three ordered edges survived FDR correction in this model. The strongest edge was \texttt{P:Patient -> primary\_ind} (\(d = 1.49\), 95\% CI [0.99, 2.16], \(q = .027\)).

For collaboration performance, the highest-AUC model was TNA with zoned learner features, with AUC = 0.765, 95\% CI [0.615, 0.902], and balanced accuracy = 0.757. Its strongest transition was \texttt{S:Info -> P:Phone} (\(d = -0.90\), \(q = .100\)), so this zoned model supports session-level separability more strongly than corrected transition-level inference. In the basic TNA representation, \texttt{information -> phone} was stronger in low collaboration-performance sessions (\(d = -1.14\), \(q = .025\)), indicating a corrected phone-related transition effect before zone labels were added. ENA performed weakly for collaboration, including one zoned-learner row below AUC = 0.50, indicating that undirected co-occurrence was not a reliable representation for this contrast.

\begin{figure}[!h]
\centering
\includegraphics[width=\linewidth]{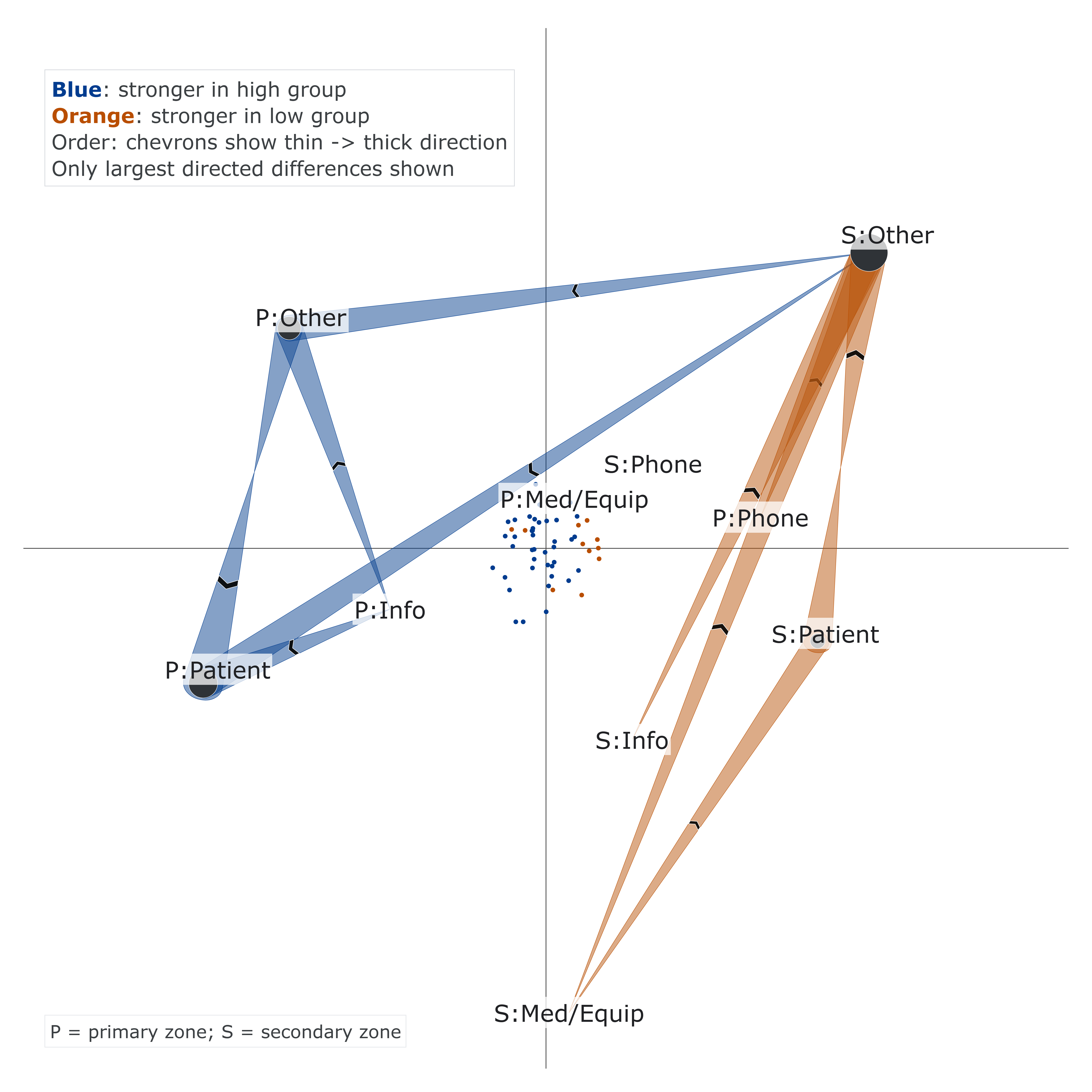}
\caption{ONA difference plot for task performance using zoned learner features. Blue edges are stronger in the high group, orange edges are stronger in the low group, and chevrons indicate ordered direction from thin to thick.}
\label{fig:task_ona}
\end{figure}

\begin{figure}[!h]
\centering
\includegraphics[width=\linewidth]{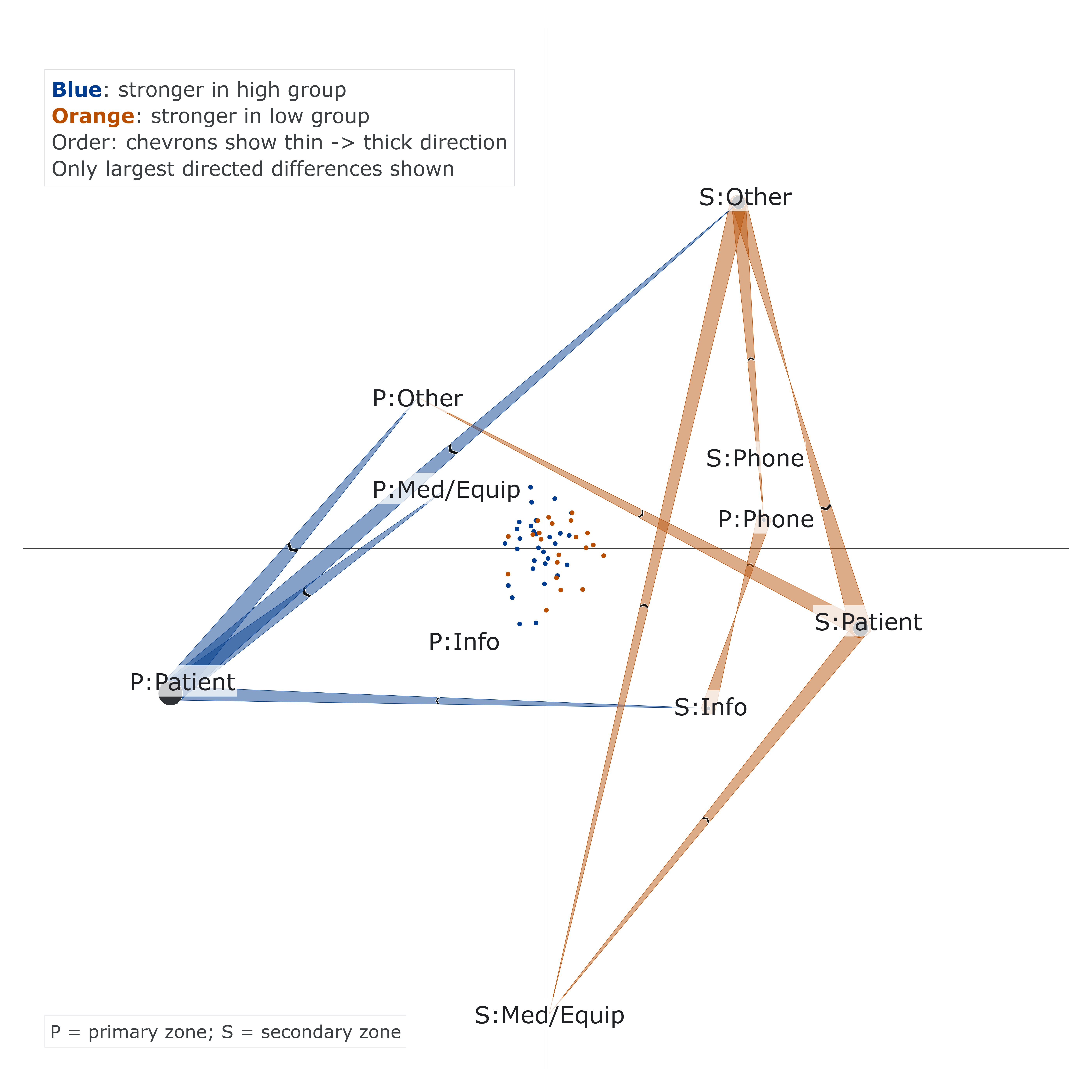}
\caption{ONA difference plot for collaboration performance using zoned learner features. Blue edges are stronger in the high group, orange edges are stronger in the low group, and chevrons indicate ordered direction from thin to thick.}
\label{fig:collab_ona}
\end{figure}

Figures~\ref{fig:task_ona} and~\ref{fig:collab_ona} visualise the ordered zoned learner models. These figures omit the co-presence nodes so that the behaviour-zone structure can be read directly.

Figure~\ref{fig:task_ona} shows stronger ordered edges into primary-zone patient interaction in high task-performance sessions, including \texttt{P:Other -> P:Patient} and \texttt{P:Info -> P:Patient}. Low task-performance sessions show stronger secondary-zone edges around \texttt{S:Other} and \texttt{S:Patient}. This matches the rate-level result in Table~\ref{tab:spatial_rates}: patient interaction differs in amount, location and ordered relation to surrounding actions.

Figure~\ref{fig:collab_ona} shows a descriptively similar pattern for collaboration performance: edges into primary-zone patient interaction appear stronger in the high group, while edges involving secondary-zone activity and phone use appear stronger in the low group. Because no collaboration ONA edge survived FDR correction, this plot is interpreted as descriptive convergence with the session-rate and classification results rather than as independent edge-level evidence.

\begin{figure}[!h]
\centering
\includegraphics[width=0.82\linewidth]{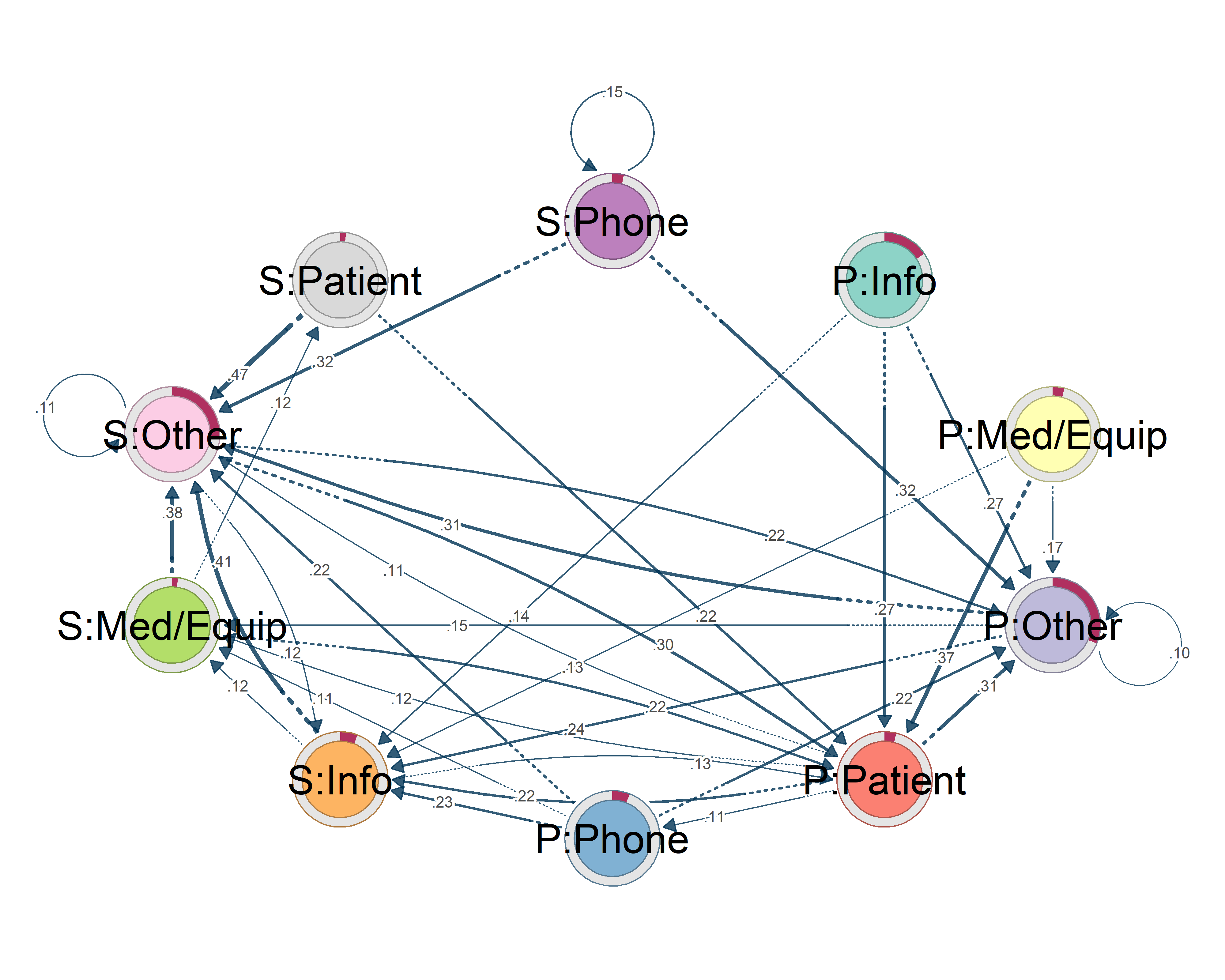}
\vspace{0.5em}
\includegraphics[width=0.82\linewidth]{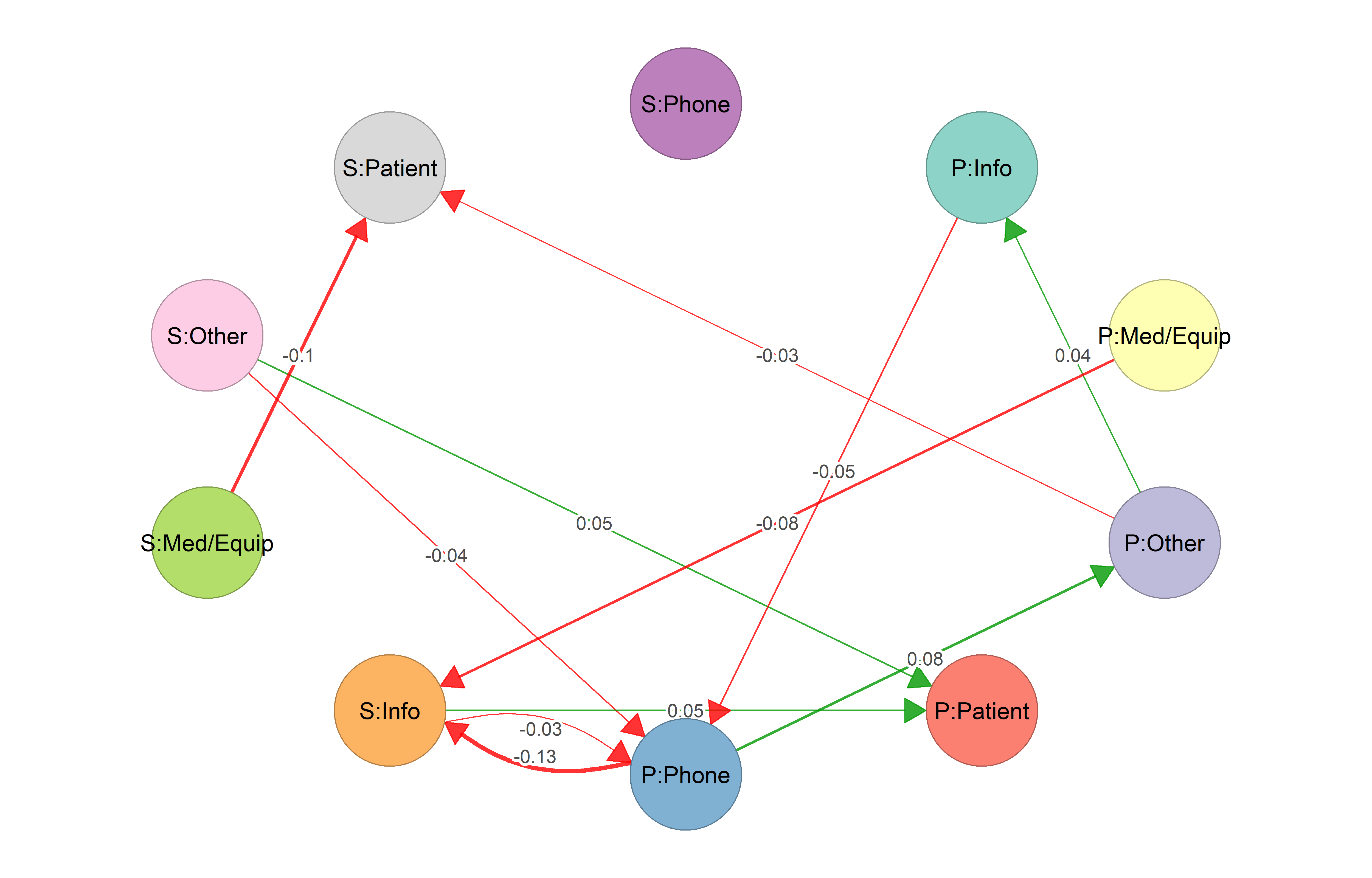}
\caption{TNA visualisation for collaboration performance using zoned learner features. The upper panel shows the global transition network after removing transitions below .10. The lower panel shows the high-low permutation contrast, with green transitions stronger in the high group, red transitions stronger in the low group and numeric labels reporting high-minus-low transition differences; this panel is descriptive because the zoned TNA transition effect did not survive FDR correction.}
\label{fig:collab_tna}
\end{figure}

Figure~\ref{fig:collab_tna} visualises the zoned learner TNA model for collaboration performance, the highest-AUC collaboration model in Table~\ref{tab:classification_results}. The upper panel shows the global transition network after filtering all transitions below .10. The lower panel shows the high-low permutation contrast. The figure should be read with the table: this zoned model separates collaboration-performance groups at the session level, but its strongest transition remains below corrected edge-level significance. The basic TNA representation, by contrast, identified one corrected phone-related transition.

DTW provided a broader temporal-shape check. DTW distances separated task-performance groups in the basic learner trace (distance difference = 0.034, \(p < .001\)) and in the zoned learner trace (distance difference = 0.056, \(p < .001\)). However, DTW classifiers were weaker than the strongest ONA and TNA models. This indicates that whole-session temporal shape differs across task-performance groups, but the more interpretable signal is carried by ordered and transition-level features.

%% file: sections/06-discussion.tex
\section{Discussion}

The results show what fixed side-view video can and cannot do in this nursing simulation archive. YOLO26 did not need to be retrained from zero when the same room was recorded from a slightly shifted side-mounted view. The detector-derived behaviour trace also contained instructor-relevant process signals, especially around phone use and patient interaction. Those signals changed meaning once location and order were added: patient interaction in the primary patient-care zone was not equivalent to patient-related activity in a secondary area. The study does not claim that computer vision can assess clinical competence. It shows how fixed-camera video can be converted into searchable behaviour-zone process evidence for simulation debriefing.

\subsection{Maintaining Fixed-Camera Analytics in Repeated Practical Teaching}

In simulation rooms, a camera may be remounted, adjusted for a later cohort or moved to avoid obstruction. The room and task ecology may remain stable, but the detector sees a different visual distribution. RQ1 tested this ordinary maintenance problem rather than a large cross-room transfer problem. Source-to-target adaptation maintained YOLO26 performance across two similar side-mounted views of the same simulation room.

The strongest practical evidence comes from the smaller 2022 target archive. Target-only training reached 0.690 mAP50, whereas full two-stage adaptation reached 0.855. For educational deployment, the largest balanced target quota of only \(N = 22\) reached 0.850 mAP50, almost matching the full two-stage result. This suggests that existing labels from a same-room source view can substantially reduce target-view annotation when a later cohort has limited labelled data. The 2021 result is more modest: full two-stage adaptation improved mAP50 from 0.815 to 0.848, while the largest balanced quota matched the target-only baseline. The annotation-saving claim is strongest for the smaller 2022 target archive, not equally strong in both directions.

The result does not imply that a model trained once will work indefinitely. A more realistic maintenance routine is to retain labelled data from previous cohorts, annotate a small balanced sample from the new view, and check whether learner-relevant behaviours remain detectable. This routine reduces full relabelling, but it still requires periodic validation. It also keeps the analytics system in a human-in-the-loop role, where technical outputs support instructional work rather than replace professional judgement \citep{gasevic2022towards, knight2017theory, dawson2019increasing, buckingham2019human}.

Per-code results matter because mean mAP50 alone could hide failure on behaviours that instructors need for debriefing. In the full two-stage setting, all five learner-relevant codes exceeded mAP50 \(\geq .70\) in both retained target views after excluding the \texttt{Sitting} control class. This includes \texttt{Medication/Equipment Interaction}, a difficult code because it involves clinical resources and treatment preparation. These detection results support timeline inspection and downstream trace analysis, but they do not justify automated assessment.

The maintenance result concerns similar-view adaptation in one repeated simulation room. It does not show robustness to arbitrary camera placements, different rooms, different patient-bed layouts, or the excluded opposite-side view. The result is about maintaining low-burden classroom analytics in a stable teaching environment, not about general computer-vision robustness.

\subsection{Behaviour-Zone Evidence for Simulation Debriefing}

RQ2 and RQ3 move from detector accuracy to the educational use of the trace. The behaviour-only trace related to instructor-labelled outcomes, but unevenly. The clearest behaviour-level signal was phone use: low task-performance sessions had higher phone-use rates. Because this signal comes from detector-generated labels rather than an independent action-coded reference, it is a cue for review rather than a validated measure of attention or disengagement.

The behaviour-only counts missed a key distinction. Total patient interaction barely separated task-performance groups, even though patient-facing care is central in this scenario. The zone split exposed the difference: high task-performance sessions had more primary-zone patient interaction, while low task-performance sessions had more secondary-zone patient interaction. The aggregate patient-interaction rate was weak because the same behaviour code carried different educational meaning depending on where it occurred.

Zone annotation changed the interpretation of \texttt{Patient Interaction}. In the primary patient-care zone, the code more plausibly indexes bedside engagement and direct patient-facing activity. In the secondary zone, visually similar or semantically related patient interaction may reflect discussion away from the immediate care space, peripheral activity, delayed return to the bedside, or movement around other resources. The detector did not infer these meanings by itself. The meaning came from combining a visible behaviour code with a task-relevant room zone.

The pattern fits spatial learning analytics, which treats the room as part of the evidence structure rather than as a neutral background \citep{martinez2018physical, yan2022spatial}. It also fits simulation debriefing practice. Instructors do not usually ask only whether students interacted with the patient; they ask when students returned to the patient, whether they stayed near the bedside after a cue, whether they moved away from patient care to consult information, and how the team distributed attention across the room. A behaviour-zone trace answers these questions more directly than a behaviour-only count.

Primary-zone co-presence gives instructors a second review cue. Higher-performing sessions had more intervals in which multiple learner-action detections appeared in the patient-care zone. Such intervals can identify moments worth reviewing for teamwork, but the variable is not a direct measure of collaboration quality. It comes from zone-coincident detections, not from speech, role negotiation, mutual monitoring or clinical reasoning.

\subsection{From Searchable Moments to Process Interpretation}

Rates show how often behaviours were detected; they do not show how actions were ordered. In practical simulation, students may consult information before returning to the patient, handle medication after a clinical cue, or shift from phone use to documentation. These relations are lost when the trace is reduced to total counts.

The ordered and transition models provide a process-level extension of the behaviour-zone findings. For task performance, ONA with zoned and co-presence features produced the strongest classifier, with AUC = 0.820 and three ordered edges surviving FDR correction. This indicates that task-performance differences involved both the amount of primary-zone patient interaction and the way that activity was ordered with surrounding behaviours and co-presence states. For collaboration performance, the highest-AUC model was TNA with zoned learner features, with AUC = 0.765. Its corrected transition-level evidence was weaker than its session-level classification performance, while the basic TNA representation identified a low-collaboration information-to-phone transition.

The task-performance result is stronger than the collaboration result. For task performance, classification and edge-level inference point in the same direction. For collaboration, the trace separates high and low sessions reasonably well, but the study cannot identify individual collaboration transitions with corrected statistical confidence. Teamwork quality is not fully visible from location and action categories; it also depends on talk, timing, role clarity and shared understanding.

ONA and TNA organise video review around patterns that instructors may want to inspect. An ordered edge into primary-zone patient interaction, for example, can guide attention to how students returned to the patient after information handling or transitional movement. A phone-related transition can identify moments where attention may have shifted away from the clinical task. These models turn whole-session video into structured prompts for professional discussion \citep{shaffer2016tutorial, zhao2023analysing}.

\subsection{Interpretive Boundaries of the Detector-Derived Trace}

The corrected holdout mAP50 results support the use of YOLO26 for deriving action traces from the retained videos, but they do not mean that the full video stream has been human coded. The RQ2 and RQ3 analyses make a narrower claim: deployed detector outputs can produce process evidence that relates to instructor-labelled session outcomes. They do not claim that the detector-derived behaviour trace is a validated substitute for human behavioural coding.

A phone-related interval, a return to primary-zone patient interaction, or a transition into co-presence can help instructors locate video segments for debriefing. The same trace cannot establish what learners intended, whether a medication decision was clinically appropriate, or whether teamwork was effective without instructor interpretation. The system should present reviewable intervals, behaviour-zone timelines and ordered patterns, not automated assessment scores.

The behaviour-zone findings have the same status. They are zone-augmented process evidence for debriefing, not direct assessment evidence.

\subsection{Design Implications for MMLA in Practical Learning Spaces}

For fixed-camera MMLA in co-located practical learning, the key design implication is that action labels alone are not enough. A useful trace must preserve the coupling between behaviour, place and temporal order because instructors use that coupling when reconstructing performance during debriefing. The analyses point to three design choices.

First, plan for view maintenance. Repeated teaching spaces change gradually, and analytics pipelines need procedures for adapting to those changes. Source-view reuse plus small target-view annotation is a practical compromise between one-off model training and full yearly relabelling.

Second, choose behaviour labels that instructors can use during debriefing. The six-code taxonomy worked because the labels corresponded to debriefing-relevant categories: information handling, medication/equipment interaction, patient interaction, phone use, and other transitional activity. Labels that are visually precise but instructionally meaningless would give instructors little to review.

Third, keep place and order in the trace. \texttt{Patient Interaction} became more interpretable when split into primary and secondary zones, and more useful again when modelled as ordered activity. This representation matches the structure of practical learning: students act in task spaces, around resources, and across time.

The same design choices may transfer to other practical courses only where the room has stable zones, recurring objects and repeated scenarios. A fire-safety setting might distinguish approaching the hazard, retrieving an extinguisher and moving through an evacuation route. A laboratory setting might distinguish apparatus handling, measurement, documentation and movement between benches. The transferable element is not the nursing taxonomy itself, but the coupling of visible action, stable task zones and instructor-interpretable outcomes.

\subsection{Limitations and Future Work}

The data come from one nursing simulation classroom and two retained similar side-mounted camera views. The excluded opposite-side view would test a stronger cross-view generalisation problem. Future work should examine whether the same adaptation strategy holds across more cohorts, camera placements, room layouts and simulation scenarios.

The learning-process analysis uses 51 labelled sessions and binary session-level instructor outcomes derived from the course assessment rubric. These labels are authentic to the teaching archive, but they are coarse. They do not identify which specific moment caused a high or low judgement, nor do they distinguish different dimensions of clinical judgement beyond the task and collaboration item groupings. The available archive does not retain independent assessor-specific scores, so inter-rater agreement for the session outcomes cannot be reported. The two outcomes are also related: all low-task sessions are low-collaboration sessions, although the high-task/low-collaboration group prevents the labels from being identical. Future work should link behaviour-zone traces to finer debriefing markers, such as response to a deterioration cue, timing of medication preparation, documentation after patient assessment, or distribution of roles during a critical event.

The sample has year-composition constraints. Low-task sessions are concentrated in 2021, with only two low-task sessions in the 2022 archive. Fisher's exact tests do not reject equal year composition, but the small cells prevent reliable year-stratified network modelling. The primary-zone patient-interaction pattern should be replicated in a larger multi-cohort archive before it is treated as stable across years.

The behaviour-zone findings depend on detector-derived action labels and zone projection. The study did not include an independent human-coded action trace for the 51 sessions, nor a human-coded zoned trace. A stronger follow-up would annotate a stratified subset of video intervals with both action and zone labels, then test whether action-detection error or zone-projection error differs systematically by outcome group.

The network and sequence results depend on modelling choices. ENA depends on stanza-window definition, ONA on the ordering procedure used to derive directed edges, TNA on transition specification, and DTW on alignment assumptions \citep{shaffer2016tutorial, zhao2023analysing, sakoe2003dynamic, petitjean2011global}. Larger data archives would allow these modelling choices to be compared more systematically across window sizes, transition definitions and model families.

Future work should examine how instructors use behaviour-zone timelines: which detected moments they trust, which they ignore, how the trace changes debriefing discussion, and whether students can use the evidence for reflection. That step is needed before fixed-camera MMLA can move from technical feasibility to sustained educational use.

%% file: sections/07-conclusion.tex
\section{Conclusion}
Fixed side-view video from repeated nursing simulation can be converted into an instructor-facing behaviour-zone trace for debriefing, but the trace must be maintained, linked to room zones and interpreted by instructors. YOLO26 source-to-target adaptation maintained detection performance after a small same-room camera-position shift, with the clearest annotation-saving result in the smaller 2022 target archive. The learning-process analyses showed why detection accuracy alone is not the endpoint. Behaviour-only rates contained signal, especially around phone use, but patient interaction became more interpretable only after adding room-zone labels. Higher-performing sessions concentrated patient interaction in the primary patient-care zone, whereas lower-performing sessions showed more secondary-zone patient interaction and phone-related patterns. Ordered and transition models added temporal structure to this behaviour-zone evidence. The value of the pipeline is not automated assessment. It is the production of searchable behaviour-zone traces that help instructors locate and interpret debriefing-relevant moments.